\documentclass[review]{elsarticle}

\usepackage{hyperref}

\usepackage{empheq}

\journal{Nucl. Instr. Meth. Phys. Res.}

\begin{document}

\begin{frontmatter}

\title{Ultrafast Large Angle Beamstrahlung Monitor}

\author{S. Di Carlo\fnref{myfootnote1}}
\address{Department of Physics and Astronomy - Wayne State University, 42 W Warren Ave, Detroit, MI 48202, USA}
\fntext[myfootnote1]{salvatore.di.carlo@wayne.edu}

\author{F. Messina\fnref{myfootnote2}}
\address{Dipartimento di Fisica e Chimica - Universita degli Studi di Palermo, Via Archirafi 36, 90133, Palermo (Italy)}
\fntext[myfootnote2]{fabrizio.messina@unipa.it}

\begin{abstract}
The analysis of beamstrahlung radiation, emitted from a beam of charged particles due to the electromagnetic interaction with a second beam of charged particles, provides a diagnostic tool that can be used to monitor beam-beam collisions in a $e^{+}e^{-}$ storage ring. In this paper we show that the beamstrahlung time profile is related to the timing of the collisions and the length of the beams, and how its measurement can be used to monitor and optimize collisions at the interaction point of the SuperKEKB collider. To measure the time dependence of beamstrahlung, we describe a method based on nonlinear frequency mixing in a nonlinear crystal of beamstrahlung radiation with photons from a pulsed laser. We demonstrate that the method allows to measure and optimize the relative timing and length of the colliding bunches with 1\% accuracy. \end{abstract}
\begin{keyword}
\texttt{beamstrahlung}\sep \texttt{SuperKEKB}\sep \texttt{Belle 2}\sep \texttt{beams}\sep \texttt{timing}\sep \texttt{monitor}\sep \texttt{collisions}\sep \texttt{SFG}\sep \texttt{nonlinear crystal}\sep \texttt{up-conversion}\sep
\texttt{frequency mixing}\sep
\end{keyword}

\end{frontmatter}


\section{Introduction}
Nowadays, the particle physics organizations are following two different but complementary approaches. The energy frontier approach consists in designing a particle accelerator that is able to provide the highest possible available energy to produce new particles or discover unknown processes at very high energies. This is the Atlas and CMS approach at LHC \cite{Evans:2008zzb}. Another approach consists in working at lower energies, thereby designing the accelerator in order to optimize the production of certain well known resonances and, studying their rare decays, underline some new processes that are not contemplated within the Standard Model. The latter is the path followed by Belle II in the framework of the SuperKEKB accelerator \cite{Abe:2010gxa}. In the latter case, one must deal with rare events (i.e., events with a small cross section) which show departures from the Standard Model. The rate of events production is given by the luminosity $\mathcal{L}$ times the cross section $\sigma$ \cite{Syphers:2013mhc}: 

\begin{equation} \label{eq:eventRate}
\frac{dN}{dt} = \mathcal{L} \sigma
\end{equation}

It is clear that the success and physical outreach of the Belle II experiment depends critically on luminosity, one of the two figures of merit of the accelerator together with the energy. The new SuperKEKB storage ring aims, through the use of nano-beams, to reach the very high luminosity of $8\times10^{35}cm^{-2}s^{-1}$ \cite{Abe:2010gxa}. The nano-beams scheme, invented by Pantaleo Raimondi \cite{Raimondi:2006}, allows to reduce the longitudinal overlapping of the beams, minimize the "hourglass effect" \cite{Lee:1999pt}, and therefore increasing the luminosity \cite{Abe:2010gxa}. The possibility of reaching such a high luminosity depends upon the ability to closely monitor the size and position of the beams. At SuperKEKB, direct monitoring of the beams at the interaction point (IP) is even more precious than in previous accelerators. The beam sizes are 50-60 times smaller than at previous colliders \cite{Olive:2016xmw}, and the high crossing angle (83 mrad) introduces a novel possible way to lose luminosity, as the two beams have to simultaneously arrive at the IP. 

To monitor the beams, SuperKEKB is equipped with several pieces of instrumentation \cite{Arinaga:2012laa}. Both storage rings are equipped with beam position monitors (BPM), which are mainly derived from the KEKB original system \cite{Arinaga:2012laa}\cite{Tejima:2000ns}. The BPMs are used to monitor the position of the beam inside the beam-pipe. When a beam goes past a bending magnet, synchrotron radiation is emitted and can be used to monitor the size of the beam. At SuperKEKB there are two categories of such monitor systems: visible light monitors and X-ray monitors \cite{Arinaga:2012laa}. There are two kinds of visible light monitors: interferometers, used to measure the horizontal size of the beams ($\sigma_x$) \cite{Arinaga:2012laa}; streak cameras, used to measure the length of the beams ($\sigma_z$) \cite{Arinaga:2012laa}. X-ray monitors will be used to measure the vertical size of the beams ($\sigma_y$) \cite{Mulyani:2016lhr}. The technique used is called ''coded aperture'' and was initially developed by astronomers, with the purpose of measuring the size of stars \cite{Dicke:1968}. The beam monitor systems described above can measure the properties of the beams far from the interaction point (IP), and therefore the properties at the IP must be extrapolated through calculations. 

The large angle beamstrahlung monitor (LABM) can measure the size of the beams at the IP \cite{Bonvicini:1997cy}, and has been successfully tested during SuperKEKB Phase I. Beamstrahlung is the radiation emitted by one beam of charged particles interacting with another beam of charged particles \cite{Augustin:1978ah}. A first prototype of LABM was designed to monitor the collisions at CESR, an $e^+e^-$ storage ring located at Cornell University \cite{Detgen:1999cm}. The LABM measures the polarization and spectrum of the radiation emitted at the IP during a collision. These properties are related to the size of the beams, therefore allowing to measure them \cite{Bonvicini:1997cy}. The LABM collects the radiation using four vacuum mirrors located inside the beam pipes. The light is then extracted through vacuum windows and travels inside a series of pipes which constitute the four LABM's optical channels. Once extracted, the properties of the light are measured inside two optical boxes located outside the interaction region. 

In this paper, we want to show that, besides polarization and spectrum, there are other important properties of the beamstrahlung light that are related to beam parameters. Specifically, we want to study the time profile of the beamstrahlung pulse that is emitted during a collision. At SuperKEKB, due to the large crossing angle, the collision timing becomes of crucial importance: if the beams do not simultaneously arrive at the IP, luminosity is lost. We will demonstrate that the time profile of the beamstrahlung pulse can be exploited to extract fundamental information about the collision timing. If the beams do not simultaneously arrive at the IP, this measurement allows adjusting the relative timing of the beams. Indeed, with respect to KEKB, the timing precision needs to improve by two orders of magnitude for a bench test comparison of 1\% luminosity loss due to timing \cite{dicarlofarhatgillard:2017}.  Beside collision timing, the method can be used to measure the longitudinal distribution (i.e, the length) of the beams, directly at the IP. 

A method to measure the longitudinal distribution of charged beams was proposed and tested at the Advanced Light Source (ALS) at Lawrence Berkeley National Laboratory \cite{Beche:2004}. When a beam goes past a bending magnet, the time profile of the synchrotron radiation emitted is measured, and this is directly related to the longitudinal distribution of the radiating beam \cite{Beche:2004}. The method is based on frequency mixing the radiation with photons from a femtosecond laser. Provided that certain conditions are satisfied, when photons from the two sources simultaneously enter a nonlinear crystal, there is a finite probability that radiation photons are upconverted to higher-energy photons within the pulse duration of the femtosecond laser \cite{Shah:1988}. Therefore, using an ultrafast pulsed laser with pulsewidth much smaller than that of the radiation, the latter one can be sampled by the former, thereby allowing to reconstruct the time dependence of the radiation pulse.

We propose an analog experimental method to measure the time profile of the beamstrahlung light emitted at the IP, by adding a new optical box exploiting the existing LABM optical channels. The novelty is that our method allows to estimate collision timing to 1\% of the length of the beams, by exploiting its relation to the asymmetry of the beamstrahlung pulse, which will be proved in the paper. Because of the importance of precisely estimating collision timing at the IP, the strategy we propose improves beam monitoring methods currently available at SuperKEKB. As a bonus, our method can measure the length of the beams directly at the IP, as we will show that the length of the radiating beam can be mapped into the beamstrahlung pulsewidth. As we will see in the next sections, the method allows to measure the length of the radiating beam with 1\% precision, which is much better than the 5\% precision allowed by streak cameras, which have a typical temporal resolution that is no better than 1 ps \cite{Scheidt:2000yg}. We remark here the importance of measurements taken at the IP: if a property of the beams is measured far from the interaction point, it has to be transported to the IP through calculation, and this introduces errors which may be significant. 

An important feature of our measurement is that it is completely shape-like. This is potentially crucial, since a shape-like measurement does not depend on absolute efficiencies, but only on the shape of the signal. In an environment such an accelerator, where data are prone to extreme noise, shape-like measurements of high precision are an advantage. We are also aware that, for a new accelerator like SuperKEKB, multiple measurements are necessary as a feedback and also to understand the dynamics of the beams.

In the first part of the paper, we present the parameters of the beams at SuperKEKB and an original Monte Carlo simulation of the collision. The original calculation is needed because in the LABM the whole "magnet" is observed, while the large angle of observation (compared to $1/\gamma$) is not suitable for standard approximations used in synchrotron radiation calculations. From the simulation, we obtain the time dependence of the radiation electric field at the LABM vacuum mirrors and relate it to geometric properties of the beams. In the final part of the paper, we thoroughly introduce the experimental method defining the important quantities, showing the properties of the nonlinear crystals, and calculating the related efficiencies. Finally, we give a description of the components that would be part of an hypothetical Ultrafast LABM optical box.

\section{Calculation of beamstrahlung fields}
\label{section:calculation}
In an effort to achieve a very high luminosity, SuperKEKB aims to work with nano-beams. Indeed, the luminosity of a collider is inversely proportional to the transverse size of the beams \cite{Syphers:2013mhc}. The parameters for SuperKEKB HER (High Energy Ring) or electron beam and LER (Low Energy Ring) or positron beam are shown in Table \ref{table:1} \cite{Olive:2016xmw}.

\begin{table}[h!]
\centering
\begin{center}
\begin{tabular}{| l | l | l | l | l | l |} 
\hline
$Beam$ & $E(GeV)$ & $N$ & $\sigma_{x} (m)$ & $\sigma_{y} (m)$ & $\sigma_{z} (m)$ \\ 
\hline
$LER(e^{+})$ & 4.0 & $9.04\times10^{10}$ & $10\times10^{-6}$ & $48\times10^{-9}$ & $6.0\times10^{-3}$ \\ 
\hline
$HER(e^{-})$ & 7.0 & $6.53\times10^{10}$ & $11\times10^{-6}$ & $62\times10^{-9}$ & $5.0\times10^{-3}$ \\ 
\hline
\end{tabular}
\end{center}
\caption{Beams parameters at SuperKEKB \cite{Olive:2016xmw}. E is the energy of the beam, N is the number of particles per bunch. $\sigma_{x}$, $\sigma_{y}$, $\sigma_{z}$ are the sizes of beam, as described in Figure \ref{fig:collision}.}
\label{table:1}
\end{table}

\begin{figure}[h]
    \centering
    \includegraphics[width=1\textwidth]{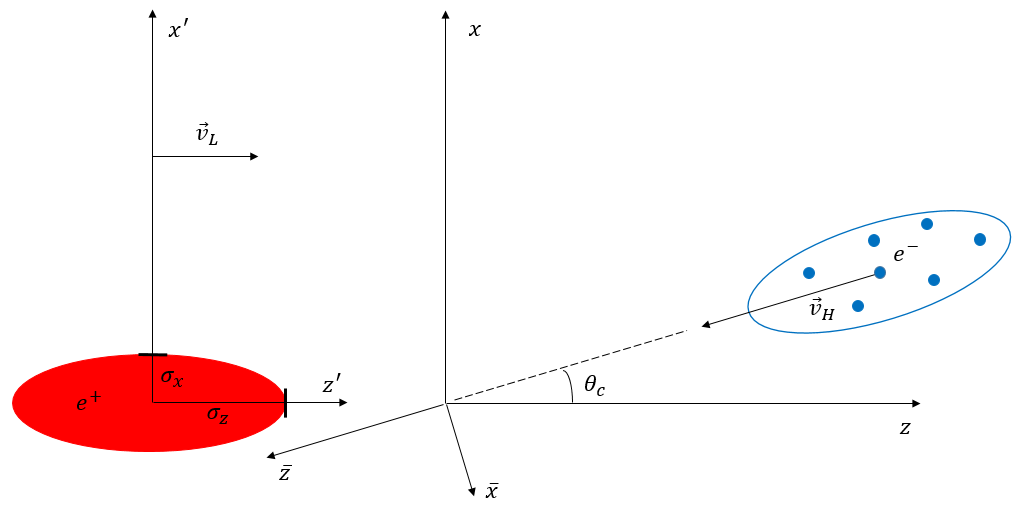}
    \caption{Geometry of the collision: the beams collide at a crossing angle of $83 \, mrad$. The electron and positron beams are organized in bunches, separated by $4 \,ns$, thereby allowing a collision frequency of 250 MhZ. The sizes of the beams are indicated as $\sigma_{x}$, $\sigma_{y}$, $\sigma_{z}$. The reference systems here defined are used in the calculation of the beamstrahlung fields.}
    \label{fig:collision}
\end{figure}

As shown in Figure \ref{fig:collision}, at the IP the collision takes place at a crossing angle $\theta_{c} = 83 \, mrad$. Every bunch is separated from the next one by $4 \,ns$, thereby allowing a collision frequency of 250 MhZ. In the collision, the electromagnetic interaction between the charged beams produces the emission of radiation, called beamstrahlung. Due to the relativistic velocities, the beamstrahlung is mostly emitted in the forward direction of motion of the beams. The two directions of motion, at the IP, are called the Oho direction for the electron beam and the Nikko direction for the positron beam. 

\medskip

We proceed now to calculate the time dependence of the beamstrahlung emitted by one bunch of electrons colliding with a bunch of positrons. The beams travel in the $x$-$z$ plane (see Figure \ref{fig:collision}). The beams are Gaussian, with sizes $\sigma_x, \sigma_y, \sigma_z$, and travel at a crossing angle $\theta_{c}$ respect to each other. The collision takes place in the origin of the $(x,y,z,t)$ reference frame, which we will call the LAB frame. We will consider the beamstrahlung emitted by one electron interacting with one beam of positrons (Figure \ref{fig:collision}). The positron beam moves with velocity $\mathbf{v_{L}}$, while the electron moves with velocity $\mathbf{v_{H}}$. The starting point is the electric field produced by the positron beam in its rest frame. The electrostatic potential $U$ in the rest frame $(x',y',z',t')$ of a charged Gaussian beam has been calculated as \cite{Kheifets:1976}, and can be used to easily obtain the electric field components generated by the beam:

\medskip

\begin{equation} \label{potential}
    U(x',y',z')=\frac{1}{4\pi\epsilon_0}
    \frac{Ne}{\sqrt{\pi}}
    \int_{0}^{+\infty} dq \frac{exp{(-\frac{x'^2}{a'^2+q}-\frac{y'^2}{b'^2+q}-\frac{z'^2}{d'^2+q})}}{\sqrt{(a'^2+q)(b'^2+q)(d'^2+q)}}
\end{equation}

\begin{equation} \label{Ex}
\resizebox{.9\hsize}{!}{$
E'_x(x',y',z') = -\frac{\partial U}{\partial x'} = \frac{1}{4\pi\epsilon_0}
\frac{2Ne}{\sqrt{\pi}}
\int_{0}^{+\infty} dq \frac{x'exp{(-\frac{x'^2}{a'^2+q}-\frac{y'^2}{b'^2+q}-\frac{z'^2}{d'^2+q})}}{(a'^2+q)\sqrt{(a'^2+q)(b'^2+q)(d'^2+q)}}
$}
\end{equation}

\begin{equation} \label{Ey}
\resizebox{.9\hsize}{!}{$
E'_y(x',y',z') = -\frac{\partial U}{\partial y'} = \frac{1}{4\pi\epsilon_0}
\frac{2Ne}{\sqrt{\pi}}
\int_{0}^{+\infty} dq \frac{y'exp{(-\frac{x'^2}{a'^2+q}-\frac{y'^2}{b'^2+q}-\frac{z'^2}{d'^2+q})}}{(b'^2+q)\sqrt{(a'^2+q)(b'^2+q)(d'^2+q)}}
$}
\end{equation}

\begin{equation} \label{Ez}
\resizebox{.9\hsize}{!}{$
E'_z(x',y',z') = -\frac{\partial U}{\partial z'} = \frac{1}{4\pi\epsilon_0}
\frac{2Ne}{\sqrt{\pi}}
\int_{0}^{+\infty} dq \frac{z'exp{(-\frac{x'^2}{a'^2+q}-\frac{y'^2}{b'^2+q}-\frac{z'^2}{d'^2+q})}}{(d'^2+q)\sqrt{(a'^2+q)(b'^2+q)(d'^2+q)}}
$}
\end{equation}

where $a'$, $b'$, and $d'$ are $\sqrt{2}$ times the standard deviations of the beam ($\sigma'_x, \sigma'_y, \sigma'_z$) and all the primed quantities are calculated in the rest frame of the beam. Now we can move to the LAB frame using the appropriate Lorentz transformations of coordinates and fields \cite{Jackson:1998nia}:

\begin{alignat}{2}
  & \begin{dcases}
  x' = x \\
  y' = y \\
  z' = \gamma_{L}(z-v_{L}t)\\
  t' = \gamma_{L}(t-\frac{v_{L}}{c^2}z)
  \end{dcases}
    & \qquad & \begin{dcases}
 \mathbf{E_\parallel} = \mathbf{E'_\parallel} \\
 \mathbf{B_\parallel} = \mathbf{B'_\parallel} \\
 \mathbf{E_\perp} = \gamma_{L}(\mathbf{E'_\perp}-\mathbf{v_{L}}\times\mathbf{B'}) \\
 \mathbf{B_\perp} = \gamma_{L}(\mathbf{B'_\perp}+\frac{1}{c^2}\mathbf{v_{L}}\times\mathbf{E'}) 
  \end{dcases}
\end{alignat}

Considering that $\mathbf{B' = 0}$, the transformation greatly simplify, and the components of the electric and magnetic fields obtained are listed below.

\begin{equation} \label{ExLAB}
\resizebox{.9\hsize}{!}{$
E_x(x,y,z,t) = \gamma_{L}
\frac{1}{4\pi\epsilon_0}
\frac{2Ne}{\sqrt{\pi}}
\int_{0}^{+\infty} dq \frac{x\,exp{(-\frac{x^2}{a^2+q}-\frac{y^2}{b^2+q}-\frac{(\gamma_{L}(z-v_{L}t))^2}{(\gamma_{L} d)^2+q})}}{(a^2+q)\sqrt{(a^2+q)(b^2+q)((\gamma_{L} d)^2+q)}}
$}
\end{equation}

\medskip

\begin{equation} \label{EyLAB}
\resizebox{.9\hsize}{!}{$
E_y(x,y,z,t) = \gamma_{L}
\frac{1}{4\pi\epsilon_0}
\frac{2Ne}{\sqrt{\pi}}
\int_{0}^{+\infty} dq \frac{y\,exp{(-\frac{x^2}{a^2+q}-\frac{y^2}{b^2+q}-\frac{(\gamma_{L}(z-v_{L}t))^2}{(\gamma_{L} d)^2+q})}}{(b^2+q)\sqrt{(a^2+q)(b^2+q)((\gamma_{L} d)^2+q)}}
$}
\end{equation}

\medskip

\begin{equation} \label{EzLAB}
\resizebox{.9\hsize}{!}{$
E_z(x,y,z,t) = \frac{1}{4\pi\epsilon_0}
\frac{2Ne}{\sqrt{\pi}}
\int_{0}^{+\infty} dq \frac{(\gamma_{L}(z-v_{L}t))\,exp{(-\frac{x^2}{a^2+q}-\frac{y^2}{b^2+q}-\frac{(\gamma_{L}(z-v_{L}t))^2}{(\gamma_{L} d)^2+q})}}{((\gamma_{L} d)^2+q)\sqrt{(a^2+q)(b^2+q)((\gamma_{L} d)^2+q)}}
$}
\end{equation}

\medskip

\begin{equation} \label{B}
\mathbf{B}(x,y,z,t) = (-\frac{v_{L}}{c^2} E_y , \frac{v_{L}}{c^2} E_x , 0)
\end{equation}

In the above formulas the gamma factor is given by $\gamma_{L}=\frac{1}{\sqrt{1-\frac{v^2_{L}}{c^2}}}$. We are now ready to calculate the Lorentz force acting on the electron. The electron travels towards the origin with velocity 

\begin{equation} \label{electronVelocity}
\mathbf{v_{H}} = (-v_{H} \sin\theta_{c} , 0 , -v_{H} \cos\theta_{c})
\end{equation}

The Lorentz force acting on the electron is

\begin{equation} \label{lorentzForce}
\mathbf{F} = -e(\mathbf{E}+\mathbf{v_{H}}\times\mathbf{B})
\end{equation}

and therefore, in our case we obtain:

\begin{empheq}[left=\empheqlbrace]{align} \label{lorentzForceComponents}
&F_{x} = -e (1 + \frac{v_{H}v_{L}}{c^2}\cos\theta_{c})E_x \\
&F_{y} = -e (1 + \frac{v_{H}v_{L}}{c^2}\cos\theta_{c})E_y \\
&F_{z} = -e (E_z - \frac{v_{H}v_{L}}{c^2}\sin\theta_{c} E_x)
\end{empheq}

At SuperKEKB, the beamstrahlung is collected by four LABM vacuum mirrors, located a few meters from the IP, in the forward propagating direction of the beam. There are  two mirrors per side, one on the top (called Up) and the other on the bottom (called Down) of the beam pipe. Therefore, it is convenient to rotate the reference system from $(x,y,z)$ to the $(\bar{x},\bar{y},\bar{z})$ coordinates (see Figure \ref{fig:collision}). The rotation is done in such a way to set the z direction as the direction of flight of the electron beam, the y direction is unchanged, and the x direction is consequently given by the right hand rule. The mirror coordinates can be expressed as:  

\begin{equation} \label{eq:mirrorPosition}
\mathbf{r} = D \hat{\bar{\mathbf{z}}} \pm D \tan{\theta}\hat{\bar{\mathbf{y}}}
\end{equation}

where D is the distance of the mirrors from the IP, $\theta$ is the elevation angle from the beam pipe axis, and the $\pm$ sign refers to the Up and Down mirrors respectively. These quantities are given in Table \ref{table:vacuumMirrorPosition}, and the elevation angle in Eq. \ref{eq:mirrorPosition} can be calculated as $\theta=(\theta_{min}+\theta_{max})/2$. 

\begin{table}[h!]
\centering
\begin{center}
\begin{tabular}{| l | l | l | l |} 
\hline
Mirror & Distance from IP (m) & $\theta_{min} (mrad)$ & $\theta_{max} (mrad)$\\ 
\hline
Oho Down & 4.51 & 8.43 & 8.87\\ 
\hline
Oho Up & 4.57 & 8.32 & 8.76\\ 
\hline
Nikko Down & 4.77 & 7.97 & 9.39\\ 
\hline
Nikko Up & 4.70 & 8.08 & 8.50\\ 
\hline
\end{tabular}
\end{center}
\caption{Vacuum mirrors positions for beamstrahlung at SuperKEKB.}
\label{table:vacuumMirrorPosition}
\end{table}

In order to move from the $(x,y,z)$ to the $(\bar{x},\bar{y},\bar{z})$ reference system, we use the following transformations:

\begin{empheq}[left=\empheqlbrace]{align} \label{eq:coordinateTransformation}
&\hat{\mathbf{x}} = -\hat{\bar{\mathbf{x}}} \cos{\theta_{c}} -\hat{\bar{\mathbf{z}}} \sin{\theta_{c}}\\
&\hat{\mathbf{y}} = \hat{\bar{\mathbf{y}}} \\
&\hat{\mathbf{z}} = \hat{\bar{\mathbf{x}}} \sin{\theta_{c}} -\hat{\bar{\mathbf{z}}} \cos{\theta_{c}}
\end{empheq}

from which we obtain:

\begin{empheq}[left=\empheqlbrace]{align} \label{eq:lorentzForceNewFrame}
&F_{\bar{x}} = - F_{x}\cos{\theta_c} + F_{z}\sin{\theta_c}\\
&F_{\bar{y}} =  F_{y} \\
&F_{\bar{z}} = - F_{x}\sin{\theta_c} - F_{z}\cos{\theta_c}
\end{empheq}

In relativistic mechanics, the force is related to the acceleration through the following relation \cite{Landau:1971vol2}:

\begin{equation} \label{relativisticForceAcceleration}
\mathbf{F} = \gamma_{H}^{3} m \mathbf{a_{\parallel }} + \gamma_{H} m \mathbf{a_{\bot }}
\end{equation}

where $\mathbf{a_{\parallel }}$ and $\mathbf{a_{\bot }}$ are the components of the acceleration that are parallel and perpendicular to the velocity of the electron respectively. Incidentally, we notice that the component parallel to the velocity of the electron will be strongly suppressed. Equation \ref{relativisticForceAcceleration} can be inverted to obtain the acceleration as a function of the force:

\begin{equation} \label{electronAcceleration}
\mathbf{a_{H}} = \frac{1}{m\gamma_{H}} \bigg(\mathbf{F} - \frac{\mathbf{v}_{H}\cdot \mathbf{F}}{c^{2}}\mathbf{v_{H}}\bigg)
\end{equation}

where m is the mass of electron. For the case at hand, the acceleration can be expressed in components as 

\begin{empheq}[left=\empheqlbrace]{align} \label{eq:accelerationComponents}
&a_{\bar{x}} = \frac{F_{\bar{x}}}{m\gamma_{H}}\\
&a_{\bar{y}} = \frac{F_{\bar{y}}}{m\gamma_{H}} \\
&a_{\bar{z}} = \frac{F_{\bar{z}}}{m\gamma_{H}^{3}} \label{eq:accelerationZ}
\end{empheq}

and therefore, we notice that the $\bar{z}$ component of the acceleration will be strongly suppressed because of the extra $\gamma_{H}^{2}$ factor in the denominator. A direct consequence is that the $\bar{z}$ component of the radiation electric field at the mirror will be strongly suppressed as well. Knowing the acceleration of the electron, it is possible to calculate the radiation field at the position of the observer \cite{Jackson:1998nia}:

\begin{equation} \label{radiationField}
\mathbf{E_{rad}}(\mathbf{r},t) = \frac{-e}{4\pi\epsilon_0 c}\bigg[\frac{\mathbf{n} \times [\mathbf{(n-\beta)}\times\mathbf{\dot{\beta}}]}{R(1-\mathbf{\beta}\cdot \mathbf{n})^3}\bigg]_{ret}
\end{equation}

where, having defined $\mathbf{w}$ as the position of the electron and $\mathbf{r}$ as the position of the observer, $\mathbf{R=r-w}$, $\mathbf{n}=R/\mathbf{R}$, $\mathbf{\beta}=\mathbf{v_{H}}/c$ and $\mathbf{\dot{\beta}}=\mathbf{a_{H}}/c$. All the quantities in the squared brackets must be calculated at the retarded time $t-R/c$. 

Having obtained the radiation electric field as a function of the time, the electric field as a function of the frequency is obtained by means of a Fourier Transform \cite{Hofmann:2004zk}:

\begin{equation} \label{eq:fourierTransform}
\mathbf{\tilde{E}_{rad}}(\mathbf{r},\omega) = \frac{1}{\sqrt{2\pi}} \int_{-\infty}^\infty \mathbf{E_{rad}}(\mathbf{r},t) e^{-i \omega t} dt
\end{equation}

Finally, the power received by the observer is given by:

\begin{equation} \label{powerMirror}
\frac{dU}{dt} = c \epsilon_{0} \mathbf{E_{rad}^2} (\mathbf{n} \cdot \mathbf{A})
\end{equation}

where $\mathbf{A}$ is the small area of the mirror that receives the beamstrahlung. In our case, the mirrors have a surface of $2.0 \times 2.8 \,mm^{2}$, and are inclined by $45^{\circ}$ with respect to the axis of the beam pipe. That makes the effective area seen from the IP equivalent to that of a $2.0 \times 2.0 \,mm^{2}$ mirror. The equations obtained in this section are used in the beamstrahlung simulation discussed in the next section.

\section{Simulation and results}
We simulate by a Monte Carlo method the collision of a Gaussian beam of positrons with a Gaussian beam of electrons. We assume that the beams are rigid, meaning that the velocity of the particles are unchanged by the interaction during the collision. For the purpose of explaining the method, let us consider the electron beam as the radiating beam, and the positron beam as the one that provides the bending force, or target beam. The Monte Carlo simulates an electron colliding with the positron beam.  Every electron will be accelerated in the collision, and therefore radiate according to the laws of classical electrodynamics \cite{Jackson:1998nia} that are given in Section \ref{section:calculation}. The result of interest is the radiation electric field calculated at the position of the four LABM mirrors as a function of the time. The simulation is then repeated for 10,000 electrons randomly distributed, according to the Gaussian distributions, within the radiating beam. Finally, the results are rescaled to take in account the nominal number of electrons present in SuperKEKB beams. The total radiation will simply be the incoherent sum of single-electron contributions. 

In this section, the $(\bar{x},\bar{y},\bar{z})$ coordinate system defined in section \ref{section:calculation} will be referred to as $(x,y,z)$ to simplify the notation. In this reference system, the z component of the electric field is strongly suppressed (see Section \ref{section:calculation}) and therefore will be neglected in the following. Therefore, here and throughout this work, we will only show the results for the x and y polarizations of the radiation electric field calculated at the position of the LABM mirrors. Of course, the same simulation can be used to simulate the case when the positron beam is the radiating beam and the electron beam is the target beam. In the following, we will present the result for just one of the four LABM mirrors, namely Nikko Down, which receives Beamstrahlung emitted by the positron beam.

\subsection{Energy spectrum}
\begin{figure}[h]
    \centering
    \includegraphics[width=1\textwidth]{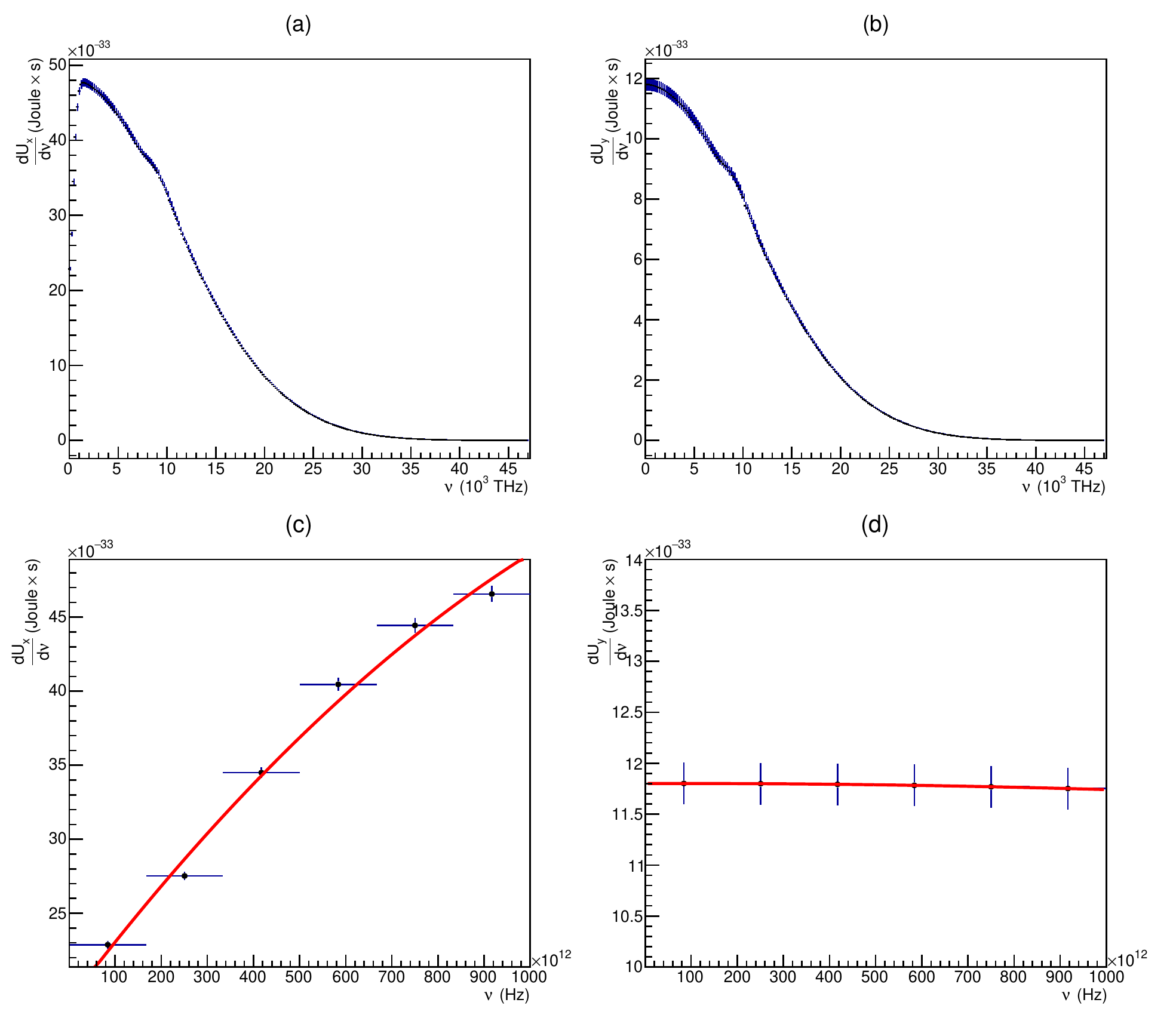}
    \caption{(a)Energy spectrum for the $x$ polarized beamstrahlung arriving at the mirror. (b)Energy spectrum for the $y$ polarized beamstrahlung arriving at the mirror. (c)Polynomial fit of the $x$ polarized energy spectrum for small frequencies. (d)Polynomial fit of the $y$ polarized energy spectrum for small frequencies. }
    \label{fig:spectrum}
\end{figure}

Fourier transforming the beamstrahlung electric field from time to frequency domain, our Monte Carlo simulation allowed us to calculate the energy spectrum of the $x$ and $y$ polarization at the mirrors. The energy spectra for $x$ polarization and $y$ polarization at the mirror are given in Figure \ref{fig:spectrum}.

The two polarizations show a different behavior at small frequencies. The spectrum for low frequencies, between 1 and 1000 THz, is shown in Figure \ref{fig:spectrum} (c,d). The spectrum for $x$ polarization increases at low frequencies, reaches a peak, and then decreases. For the y polarization, instead, we notice that the energy spectrum is flat at low frequencies and then decreases. Of course, in practice the beamstrahlung can be measured only for a limited subset of frequencies. In this paper, we will focus on the visible spectrum (430-770 THz), since it is the part of spectrum which is of concern for our purposes. A quadratic fit of the data obtained by the simulation was used to gain a higher detail for the spectrum at small frequencies. The latter is shown in Figure \ref{fig:spectrum}.

Finally, we have calculated the number of visible photons per pulse that arrive at the mirror. In table \ref{table:photonsAtmirror}, we show the total visible energy per collision at the mirror, the corresponding number of photons, and the size of the spot where the light is collected.

\begin{table}[h!]
\centering
\tiny
\begin{center}
\begin{tabular}{ | l | l | l | l | l | l | l |} 
\hline
Mirror & $U_{x}$ $(10^{-18} J)$ & $U_{y}$ $(10^{-18} J)$ & $n_{VIS,x}$ & $n_{VIS,y}$ & $dn_{VIS,x}/dt$ & $dn_{VIS,y}/dt$\\ 
\hline
Oho Down & 5.04 & 2.54 & 12.88 & 6.56 & $3.22\times10^{9}$ & $1.64\times10^{9}$\\ 
\hline
Oho Up & 5.12 & 2.60 & 13.08 & 6.73 & $3.27\times10^{9}$ & $1.68\times10^{9}$\\ 
\hline
Nikko Down & 12.96 & 4.02 & 33.15 & 10.39 & $8.29\times10^{9}$ & $2.60\times10^{9}$\\ 
\hline
Nikko Up & 14.03 & 3.88 & 35.97 & 10.02 & $8.99\times10^{9}$ & $2.51\times10^{9}$\\ 
\hline
\end{tabular}
\end{center}
\caption{$U_{x}$ and $U_{y}$ are the visible (430-770 ThZ) energies per pulse for the $x$ and $y$ polarizations at the vacuum mirrors. $n_{VIS,x}$ and $n_{VIS,y}$ are the corresponding number of photons. $dn_{VIS,x}/dt$ and $dn_{VIS,y}/dt$ are the photons per unit second arriving at the mirror for the $x$ and $y$ polarizations.}
\label{table:photonsAtmirror}
\end{table}

The temporal distribution of the visible photons within the pulse does not exactly follow the overall distribution. Indeed, the beamstrahlung pulsewidth for visible photons is somewhat larger than that for the total pulse. The reason is that hard photons are emitted mainly in the central part of the collision, while visible photons are emitted also in the tails and/or when beams are further apart. In the following, we will show the results obtained for photons corresponding to 600 THz.

\subsection{Pulse skewness and beam timing}
\begin{figure}[h]
    \centering
    \includegraphics[width=1\textwidth]{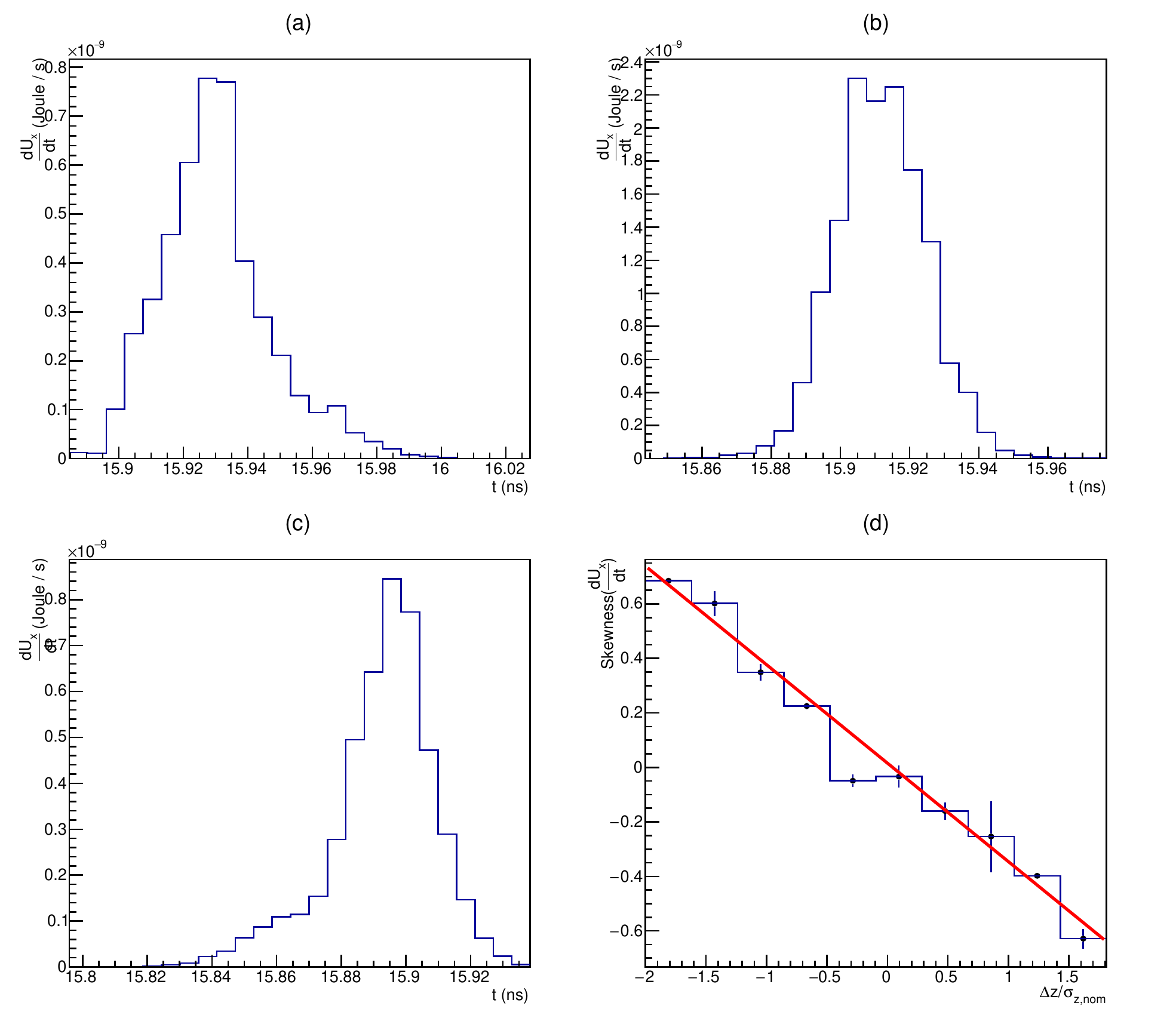}
    \caption{(a)$x$ power for delaying beam, positive skewness. (b)$x$ power for beam in time, zero skewness. (c)$x$ power for advancing beam, negative skewness. (d)Skewness of $x$ power versus delay of the radiating beam: on the horizontal axis, we have the ratio of $\Delta_{z}$ and the nominal length of the radiating beam; linear fitting in red. }
    \label{fig:skewness}
\end{figure}

The fundamental result of our simulation is that the symmetry of the beamstrahlung pulse depends on the timing of the colliding beams. In order to show this, we define $\Delta{z}$ as the distance between the centers of the two beams at the instant when the center of the target beam corresponds to the IP. The calculated skewness for the beamstrahlung pulse is shown in Figure \ref{fig:skewness} as a function of $\Delta{z}$. We see that if the radiating beam is delaying, the beamstrahlung pulse will have a positive skewness (Figure \ref{fig:skewness}-a). If the radiating beam is in time, the beamstrahlung pulse will have zero skewness (Figure \ref{fig:skewness}-b). Finally, if the radiating beam is in advance, the beamstrahlung pulse will have a negative skewness (Figure \ref{fig:skewness}-c). Therefore, measuring the skewness of the beamstrahlung pulse, it is possible to establish if the emitting beam is advanced or delayed respect to the target beam, providing a measure of the relative timing of the two beams. 

We notice that the skewness dependence for the $x$ polarization is approximately linear (Figure \ref{fig:skewness}-d). We have positive skewness for negative $\Delta{z}$, corresponding to the radiating beam arriving at the interaction point after the target beam. Conversely, we have negative skewness for positive $\Delta{z}$, corresponding to the radiating beam arriving at the interaction point before the target beam. For the $y$ polarization we have essentially zero skewness, the fluctuations due to the statistical nature of the simulation. Therefore, only the $x$ polarization component of the pulse can be used to monitor the timing of the beams.

We notice that the skewness is small for beams close to perfect timing, possibly making difficult a measure in case of very small $\Delta{z}$. However, we propose a strategy that makes a precise adjustment of the relative timing possible though the observation of the beamstrahlung skewness. Purposely changing the timing of the beams, we can move to large positive $\Delta{z}_1$ and then to large negative $\Delta{z}_2$ corresponding to the same skewness in absolute value. Finally, we can average $\Delta{z}_1$ and $\Delta{z}_2$, thereby obtaining the point of zero skewness, corresponding to beams perfectly in time.

\subsection{Pulse duration and beam length }
\begin{figure}[h]
    \centering
    \includegraphics[width=1\textwidth]{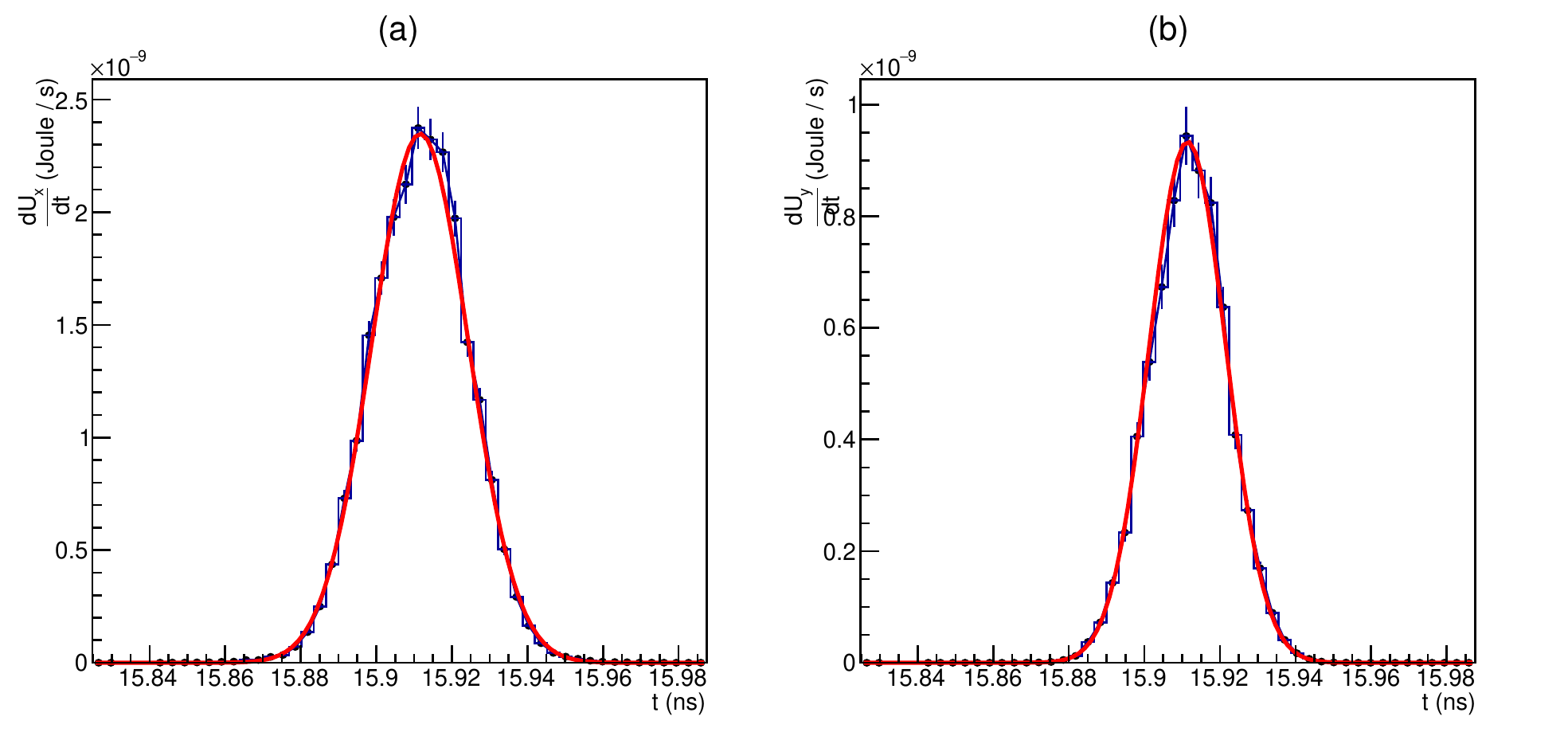}
    \caption{(a)Beamstrahlung power for $x$ polarized photons. (b)Beamstrahlung power for $y$ polarized photons. }
    \label{fig:power}
\end{figure}

The other important measure is the beamstrahlung pulse duration. We show, in Figure \ref{fig:power}, the power of the beamstrahlung pulse as a function of the time. The beamstrahlung pulse is received by the mirror about 15 ns after the beams collision, lasting for a time interval about 10 ps long (rms). 

\begin{figure}[h]
    \centering
    \includegraphics[width=1\textwidth]{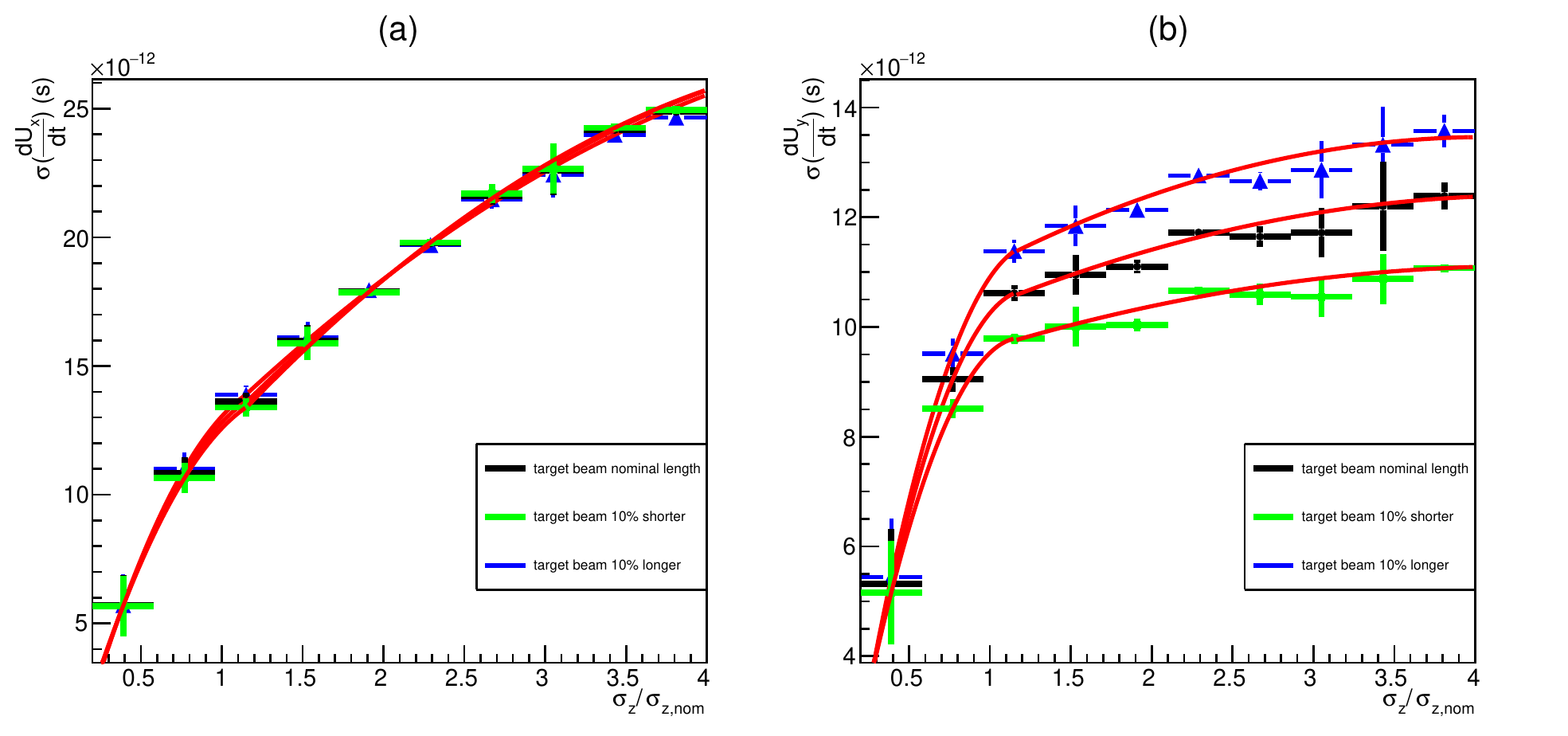}
    \caption{Pulse duration (rms) versus beam length (rms) for photons corresponding to 600 THz. (a) $x$ polarization; (b) $y$ polarization. On the horizontal axis, we have the ratio of the beam length $\sigma_{z}$ used in the simulation and the nominal one. We consider three cases: nominal conditions of the target beam (black), target beam 10\% shorter (green) and 10\% longer (blue). The result refers to photons corresponding to 600 THz. Polynomial fitting in red.}
    \label{fig:sigma}
\end{figure}

We notice that the duration of the pulse depends on the polarization, with the $x-$polarization being slightly wider. The result shown in Figure \ref{fig:power} was obtained with beams in nominal conditions. Interestingly, the  beamstrahlung pulse duration is strictly related to the length of the beam, $\sigma_{z}$. Of course, the length of the pulse depends also on the length of the target beam. We consider three cases here: nominal conditions of the target beam, target beam 10\% shorter, and 10\% longer. For each case we varied the length of the radiating beam, thereby obtaining the corresponding temporal duration (RMS) of the beamstrahlung pulse, reported in Figure \ref{fig:sigma}.

We notice that the dependence is approximately linear for a beam of length shorter or equal to the nominal value, while it tends to a plateau for a longer beam. This behavior for long beams is explained because, the target beam being much shorter, the interaction only takes place within the central part of the radiating beam. Therefore, the simulation demonstrates that by measuring the time dependence of the beamstrahlung pulse it is possible to establish the length of the radiating beam at the IP.

From what we have seen, the $x$ polarization is the richer one, since its study allows to measure both the timing of the collisions and the length of the beams. As an average situation, based on the results in Table \ref{table:photonsAtmirror} we will consider 10 visible photons per collision arriving at the mirror with $x$ polarization. In the next section, we will describe the method to measure such photons and reconstruct the beamstrahlung time dependence.

\section{Method of measurement}
Dealing with very short light pulses is a challenging task, because electronics is not able to measure pulses shorter than about 100 ps \cite{Ronzhin:2015zmc}, while streak cameras have a typical temporal resolution that is $\sim$1 ps at best \cite{Scheidt:2000yg}. To overcome these limits, physicists have developed techniques which, using femtosecond lasers in combination with nonlinear optics, allow to manipulate and measure light pulses with a temporal resolution down to a few femtoseconds~\cite{Shah:1988}. There are potentially many ultrafast techniques that would be suitable to measure the time profile of beamstrahlung pulses. For instance, one can exploit a material trasparent in the visible range, e.g. a wide band gap semiconductor or a UV-absorbing fluorescent dye, and excite it via non-degenerate two photon absorption (TPA) of visible photons from the beamstrahlung beam, arriving simultaneously to the near-IR photons from a femtosecond lasers\cite{Xue:2015}. Since TPA is only possible if the two pulses overlap in time, this phenomenon can be used to reconstruct the time profile of B pulses by scanning the delay between the two beams, and detecting either the fluorescence emitted by the excited sample (if any), or the change in transmission of the B beam caused by TPA. Any other process due to the  nonlinear interaction of the two pulses, such as so-called cross-phase modulation\cite{Lorenc2002}, may be similarly used to the same aim. Here we will focus on, and discuss in detail, a method that exploits the idea of photon upconversion, a powerful technique allowing to obtain a temporal resolution that is approximately given by the pulsewidth of the laser\cite{Shah:1988}. This method is founded on sum frequency generation of the beamstrahlung with an intense, pulsed laser beam within a nonlinear crystal. A similar approach is currently used to measure fluorescence emission with sub-picosecond time resolution\cite{B500108K}\cite{Messina:2013}, and was also used to measure the length of beams emitting synchrotron radiation while progressing through a bending magnet \cite{Beche:2004}.

In the following, the beamstrahlung pulse will be referred to as the B pulse, while the laser pulse will be referred to as the P pulse. In our discussion we will refer to a P beam of wavelength 800 nm, typical of femtosecond Ti:Sapphire lasers, while for the beamstrahlung we will consider photons of wavelength 500 nm, or equivalently a frequency of 600 THz. Nowadays P pulses are as short as few femtoseconds, therefore much shorter than the B pulsewidth, about 20 ps based on the results of the previous section. More specifically, we will consider a P laser of pulsewidth 50 fs, average energy per pulse 10 nJ, corresponding to a 0.2 MW peak power. The measurement method we propose can be shortly described as follows. The P and B pulses are sent to overlap within a crystal endowed with marked nonlinear optical properties, such as $\beta$-Barium Borate (BBO) or Lithium Iodate. Within the crystal, there is a finite probability that a sum-frequency generation process takes place, generating new photons with energies and wavevectors given by:

\begin{empheq}[left=\empheqlbrace]{align} 
&\nu_{B} + \nu_{P} =  \nu_{S}\\
&\mathbf{k_{B}} + \mathbf{k_{P}} =  \mathbf{k_{S}} 
\label{eq:phasematching1}
\end{empheq}

When P pulses are in the near-infrared and B pulses are in the visible range, the generated S photons will be in the ultraviolet (e.g. 308 nm in our example). This is tantamount to upconverting B photons to higher frequencies within the interaction time window with the P beam. In the following, the properties referred to upconverted photons will be labelled with an S. Upconverted photons are then measured and, if the delay between the two beams is changed during the measurement, one can thus use the shorter P pulse to scan the B pulse in order to reconstruct its original time profile. Of course, the synchronization of the laser with the bunch is of fundamental importance. This has already been achieved, while maintaining the laser stable and synchronized with the bunch within a resolution better than 150 fs \cite{Schulz:2010} \cite{Schulz:2015}.

\subsection{Phase-matching conditions}
Assuming collinear beams, Eq. \ref{eq:phasematching1} can be rewritten as:
\begin{equation}
\frac{n_S}{\lambda_S}=\frac{n_B}{\lambda_B}+\frac{n_P}{\lambda_P}
\label{eq:phasematching2}
\end{equation}

Because of the wavelength-dependence of the refractive indexes, this, so-called, phase-matching condition cannot be trivially satisfied, strongly limiting the efficiency of the nonlinear process. However, such a problem can be overcome by exploiting the birefringence of the nonlinear crystal, allowing to propagate waves with orientation-dependent refractive indexes. Here we will assume the use of uniaxial crystals such as BBO, namely crystals with only one symmetry axis. A wave propagating in such a crystal experiences a refractive index $n_{o}(\lambda)$ (ordinary refractive index) if its polarization is perpendicular to the optical axis. In contrast, if the polarization lies in the plane defined by the wavevector and the optical axis, the beam is called extraordinary, and the refractive index is given by:
\begin{equation}
\frac{1}{n^{2}(\theta,\lambda)}=\frac{sin^2(\theta)}{n_{e}^{2}(\lambda)}+\frac{cos^2(\theta)}{n_{o}^{2}(\lambda)}
\label{refractive}
\end{equation}
where $\theta$ is the angle between the electric field and the optical axis and $n_{e}(\lambda)$  is called the extraordinary refractive index.

While the phase matching conditions  cannot be satisfied if all waves, B, P, and S, are ordinary waves, Eq. \ref{eq:phasematching2} can be fulfilled by a suitable choice of the beam polarizations and of the polar angle $\theta$, because the latter allows to continuously tune the refractive index of the extraordinary wave through Eq. \ref{refractive}. In fact, considering an interaction where B and P beams are ordinary waves, while S is extraordinary ($O + O \to E$ interaction), the refractive indexes of the B, P, and S waves are n$_o$($\lambda_B$), n$_o$($\lambda_P$), n($\theta$,$\lambda_S$), respectively. Hence, from Eqs. \ref{refractive} and \ref{eq:phasematching2}, we obtain the following expression for the phase-matching angle:

\begin{equation} \label{eq:theta_m}
\sin^{2}{\theta_{m}} = \frac{(1/n^{2}(\theta_{m},\lambda_S))-(1/n_{o}^{2}(\lambda_{S}))}{(1/n_{e}^{2}(\lambda_{S}))-(1/n_{o}^{2}(\lambda_{S}))}
\end{equation}

where $n(\theta_{m},\lambda_S)$ is obtained by Eq. \ref{eq:phasematching2} as follows:

\begin{equation}
n_S(\theta_{m},\lambda_S)=\frac{n_o(\lambda_B)\lambda_S}{\lambda_B}+\frac{n_o(\lambda_P)\lambda_S}{\lambda_P}
\end{equation}

Provided that B and P beams are polarized as ordinary waves, upconverted photons will be efficiently generated by adjusting the polar angle $\theta$ to the value $\theta_m$. The choice of $\theta$ hence establishes the wavelength $\lambda_{B}$ of the B photon that will be efficiently upconverted. In practice, upconversion will affect photons within a narrow bandwidth of frequencies centered about $\nu_{B}$. The bandwidth can be expressed as \cite{Shah:1988}:

\begin{equation} \label{eq:spectralBandwidth}
\Delta \nu_{B} (Hz) = \frac{0.88}{L(cm)[\gamma_S(s/cm) - \gamma_B(s/cm)]}
\end{equation}

where:

\begin{equation} \label{eq:gamma_B}
\gamma_{B}=\frac{1}{c} \bigg[n_{o}(\lambda_B) - \lambda_{B}\frac{\partial n_{o}}{\partial\lambda} \bigg|_{\lambda = \lambda_{B}} \bigg]
\end{equation}

and

\begin{equation} \label{eq:gamma_S}
\gamma_{S}=\frac{1}{c} \bigg[n_{S}(\theta_{m},\lambda_S) - \lambda_{S}\frac{\partial n_{S}(\theta_{m},\lambda)}{\partial\lambda} \bigg|_{\lambda = \lambda_{S}} \bigg]
\end{equation}

\begin{figure}[h]
    \centering
    \includegraphics[width=1\textwidth]{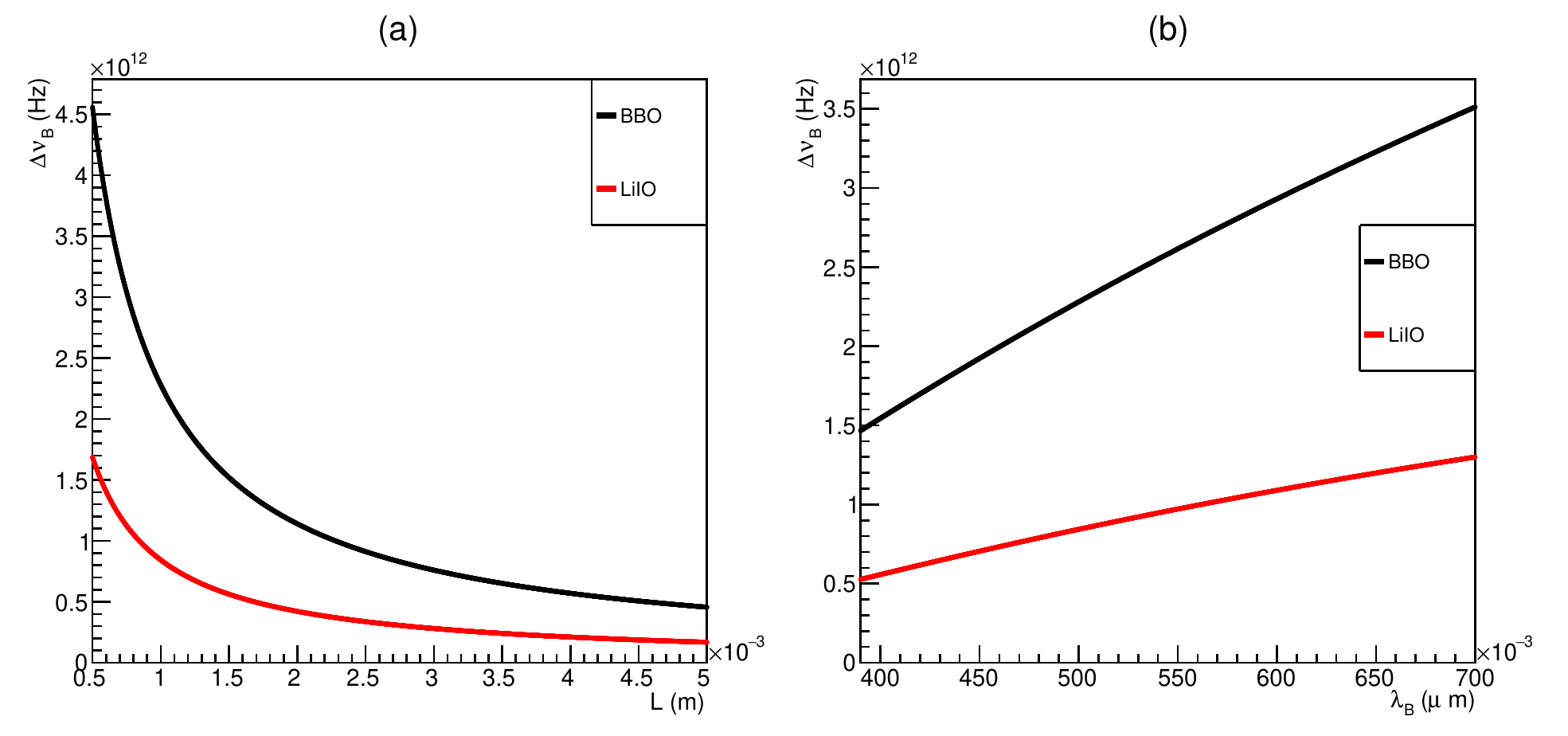}
    \caption{(a)Spectral Bandwidth vs crystal length with $\lambda_B=500\,nm$. (b)Spectral Bandwidth vs wavelength with L=1mm. BBO crystal in black; LiIO crystal in red. In the calculations, the refractive indexes of the crystals were calculated by using Sellmeier equations.}
    \label{fig:bandwidth}
\end{figure}

The bandwidth is shown in Figure \ref{fig:bandwidth} as a function the crystal length and $\lambda_B$ for two different nonlinear crystals. Its order of magnitude is about 1 THz for 1 mm-thick nonlinear crystals. Because the beamstrahlung radiation is very polycromatic, the limited spectral acceptance bandwidth of the nonlinear process will significantly reduce 
the rate of photons upconverted, and therefore it is an important parameter to take into account when estimating the efficiency of this measurement method.

In order to enhance the efficiency of the process, it is useful to focus the beams in order to increase the local intensity traversing the nonlinear crystal. However, to have upconversion, the P and B pulses incoming on the crystal must arrive within a certain solid angle of acceptance. Because it is easier to regulate the convergence of the laser than of the B beam, the most critical condition concerns the latter. Under certain conditions, the acceptance angle for the B beam is approximately given by \cite{Zernike:1973}:

\begin{equation} \label{eq:acceptanceAngle}
\Delta\phi= \frac{2.78 n_{o}(\lambda_B) \lambda_{B}}{L[1-(n_{o}(\lambda_B)\lambda_{S})/(n_{S}(\theta_{m},\lambda_S)\lambda_{B})]}
\end{equation}

\begin{figure}[h]
    \centering
    \includegraphics[width=1\textwidth]{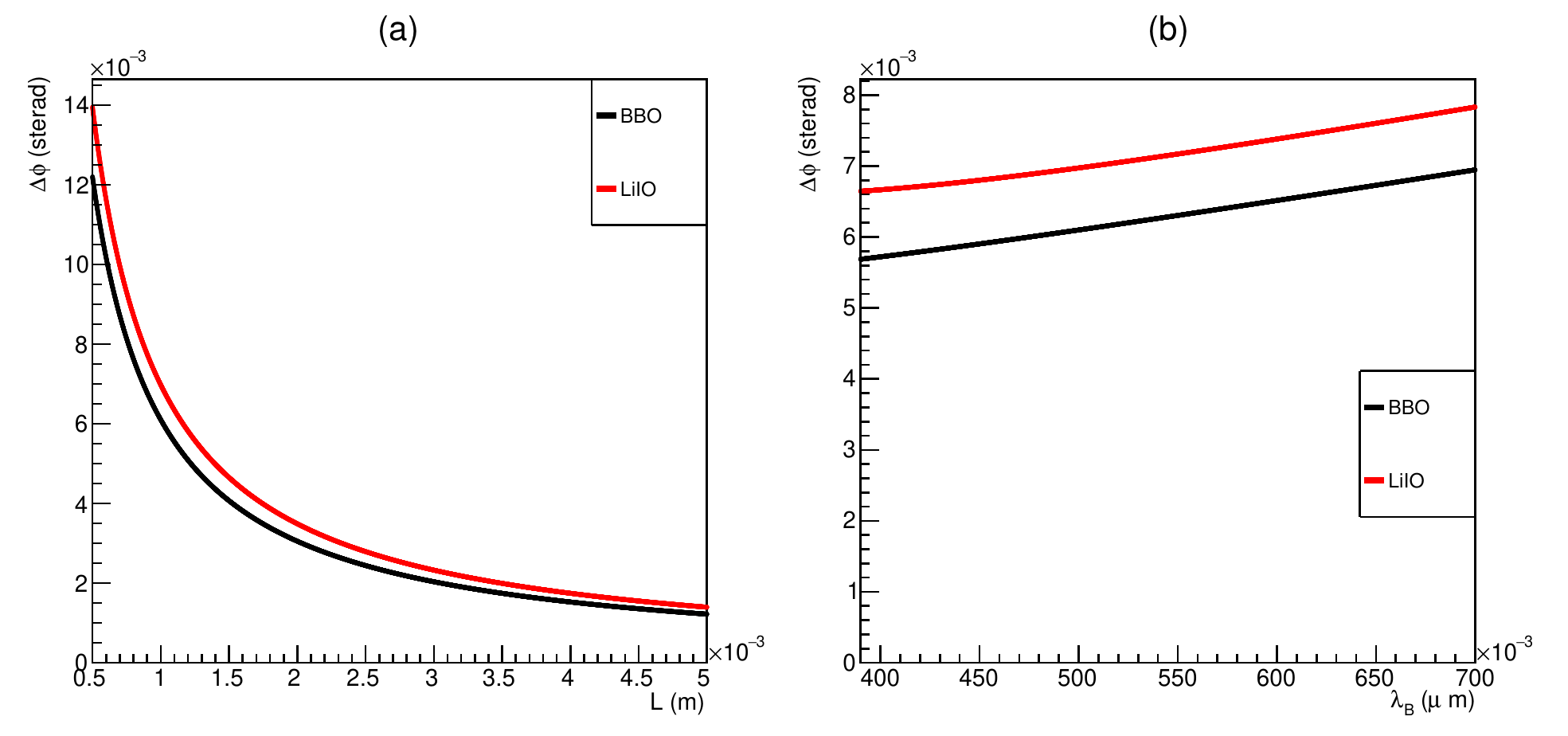}
    \caption{(a)Angle of acceptance vs crystal length with $\lambda_B=500\,nm$. (b)Angle of acceptance vs wavelength with L=1mm. BBO crystal in black; LiIO crystal in red.}
    \label{fig:acceptanceAngle}
\end{figure}

The angle of acceptance is plotted in Figure \ref{fig:acceptanceAngle}. To avoid a reduction of the upconversion rate, it is then important to focus the B beam within this solid angle. Given the expected 6 mm diameter of the B beam at the optical box, and assuming 1 mm-thick nonlinear crystals, we estimated that this condition can be fulfilled by focusing it with a converging lens (or mirror) with a focal length of about 2000 mm. In these conditions, the acceptance angle should not limit the overall conversion efficiency.

\subsection{Group Velocity Mismatch and time resolution}
Since the refractive indexes depend on the wavelength, the three pulses B, P, and S have different group velocities within the crystal. This fact can cause a temporal broadening of the pulses and therefore a deterioration of the time resolution \cite{Shah:1988}. The group velocity mismatch is given by \cite{Shah:1988}:

\begin{equation} \label{eq:deltaT}
\Delta t(s) = L(cm) \bigg[\gamma_{P}(s/cm) - \gamma_{B}(s/cm) \bigg]
\end{equation}

where

\begin{equation} \label{eq:gamma_P}
\gamma_{P}=\frac{1}{c} \bigg[n_{o}(\lambda_P) - \lambda_{P}\frac{\partial n_{o}(\lambda)}{\partial\lambda} \bigg|_{\lambda = \lambda_{P}} \bigg]
\end{equation}

\begin{figure}[h]
    \centering
    \includegraphics[width=1\textwidth]{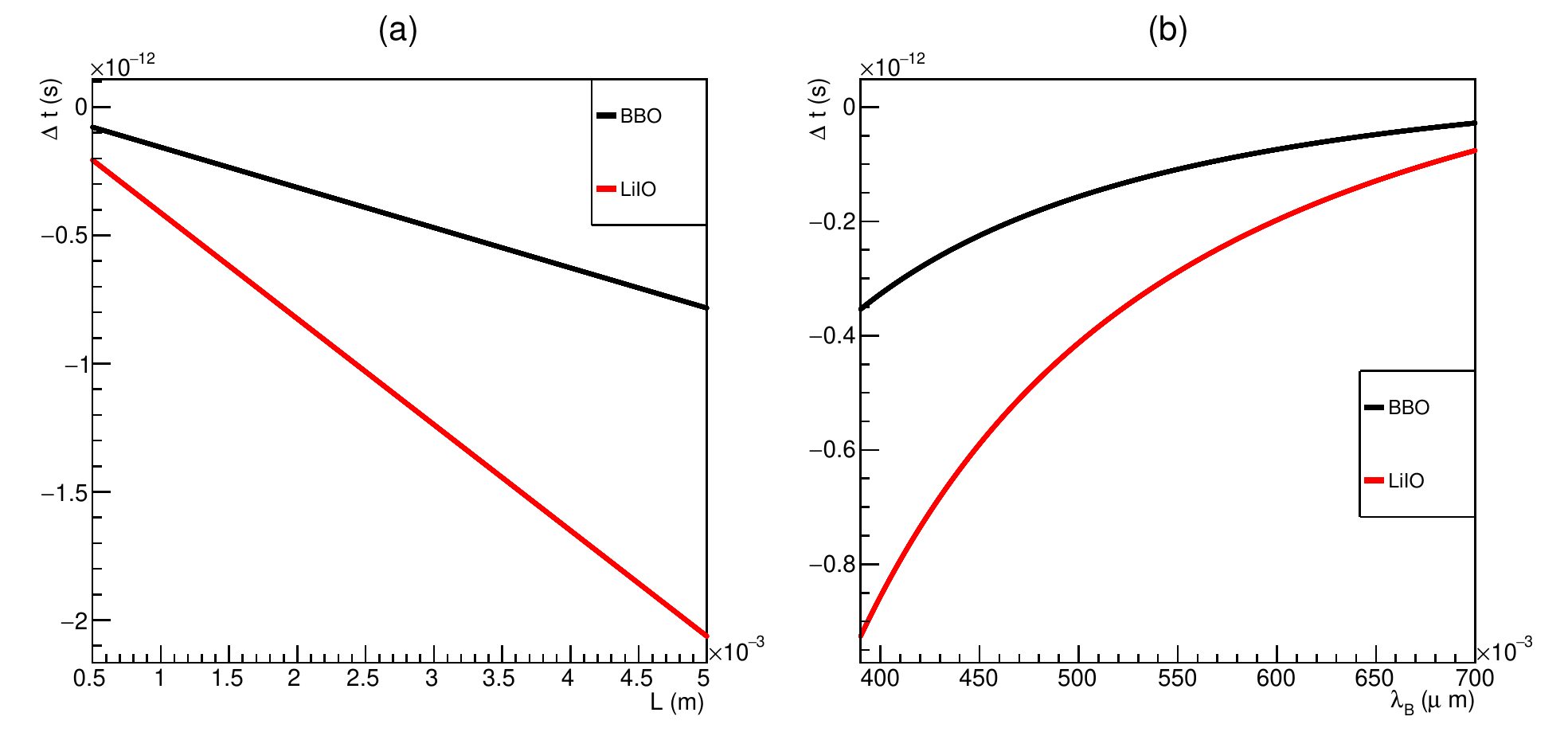}
    \caption{(a)Group velocity vs crystal length with $\lambda_B=500\,nm$. (b)Group velocity vs wavelength with L=1mm. BBO crystal in black; LiIO crystal in red.}
    \label{fig:groupVelocity}
\end{figure}

The group velocity is shown in Figure \ref{fig:groupVelocity} as a function the crystal length and $\lambda_B$. We clearly see that, if we want a resolution of at least 200 fs, we need to use a crystal no longer than 1 mm. Such a resolution is one hundredth of the total length of the B pulse, and therefore we can measure the B pulsewidth to 1\% accuracy. It directly follows that, with such a resolution, it is possible to measure the length of the radiating beam with 1\% confidence.

\subsection{Efficiency of photon upconversion}
We are now ready to calculate the rate of beamstrahlung photons upconverted. For the nonlinear crystal, we will consider a BBO of length 1 mm. The pulsed laser will have a wavelength of 800 nm, pulsewidth $2\sigma_{P}=50 fs$, average power 10 nJ, and peak power 0.2 MW. 

For the beamstrahlung we will consider photons of wavelength 500 nm. We remind the reader that the beamstrahlung pulsewidth (2 times the RMS) is about $2\sigma_{B}=20 \, ps$, there are about $n_{VIS}=10$ visible photons per pulse with $x$ polarization, and the collision frequency is $f=250 \, MHz$. We suppose to focalize the two pulses on an area $A$ of diameter $400 \, \mu m$ on the nonlinear crystal. The rate of upconverted photons will be given by:

\begin{equation} \label{eq:upconvertedRate}
\frac{dN_{up}}{dt}  \approx n_{VIS} \times \frac{f}{3} \times \frac{\Delta\nu_{B}}{\Delta_{VIS}} \times \frac{\sigma_{P}}{\sigma_{B}} \times \eta_{0}
\end{equation}

where $\Delta_{VIS}=(770-430) \, ThZ=340 \, ThZ$ and $\Delta\nu_{B}$ is the spectral bandwidth of upconversion. The collision frequency is divided by 3 because the repetition rate of a typical commercial Ti:Sapphire oscillator usually ranges around 80 MHz, which can be precisely synchronized to the third sub-harmonic of the collider (83.33 MHz),\cite{Halcyon}, and therefore we can only measure one third of beamstrahlung pulses. 

\begin{figure}[h]
    \centering
    \includegraphics[width=1\textwidth]{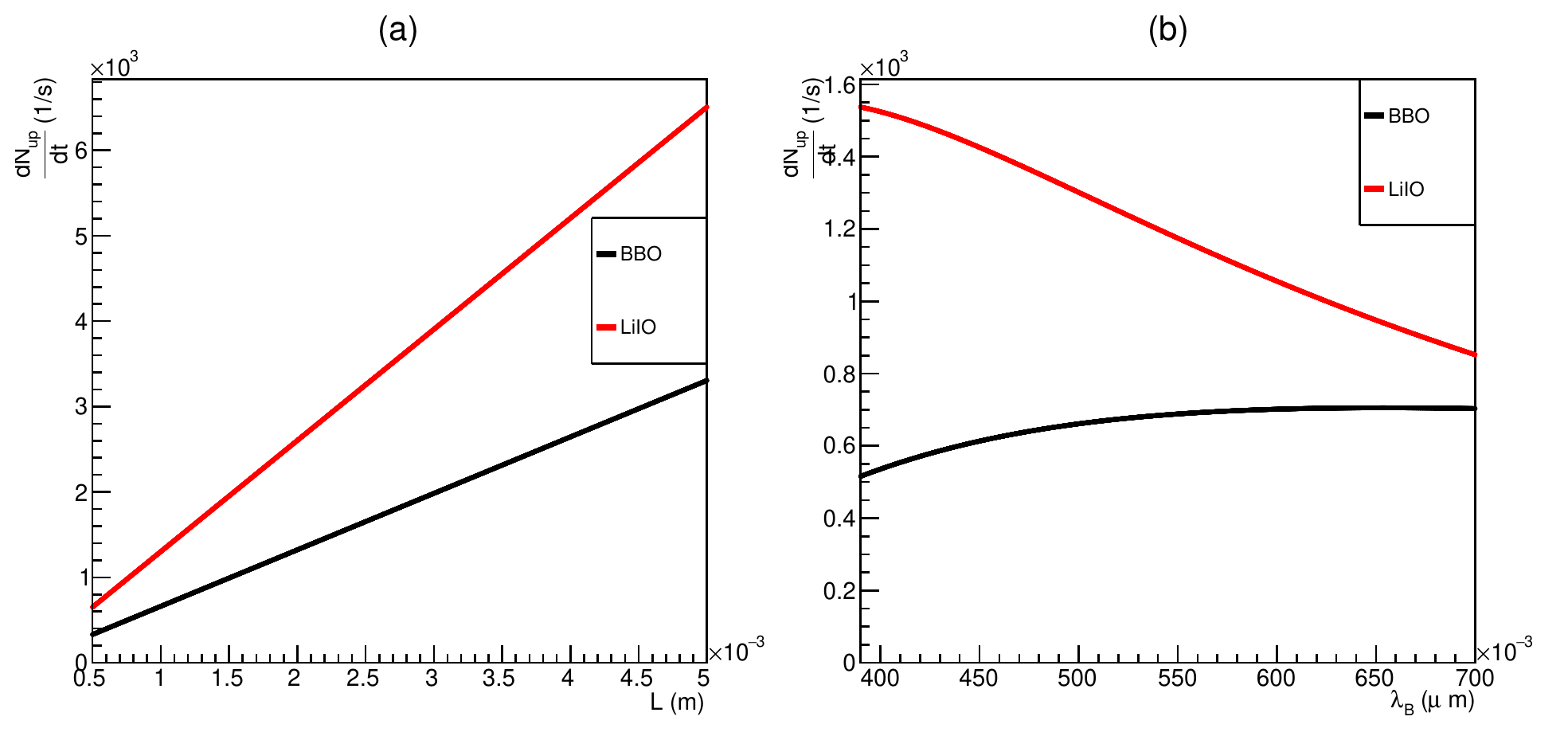}
    \caption{(a)Number of upconverted photons vs crystal length with $\lambda_B=500\,nm$. (b)Number of upconverted photons vs wavelength with L=1mm. BBO crystal in black; LiIO crystal in red.}
    \label{fig:photonsUp}
\end{figure}

The quantum efficiency of upconversion $\eta_0$, appearing in Eq. \ref{eq:upconvertedRate}, is given by \cite{Zernike:1973}\cite{Shen:1984}:

\begin{equation} \label{eq:quantumEfficiency}
\eta_{0} = \frac{2 \pi^2 d_{eff}^2(P_{P}/A)L^{2}}{\lambda_{B}\lambda_{S} n_{o}(\lambda_B) n_{o}(\lambda_P) n_{S}(\theta_{m},\lambda_S)c \epsilon_0^3}
\end{equation}

where $P_{P}$ is the peak power of the pulsed laser, $A$ is the area where the P beam is focused on the crystal (assuming the B beam is focused on an area no larger than $A$) and $d_{eff}$ the effective nonlinear coefficient of the crystal. The latter depends on the structure of the crystal and also on its orientation respect to the incoming beams. For a BBO crystal, phase-matched to upconvert 500 nm, the effective nonlinear coefficient equals $1.9$ $pm/V$. From this value, and using the above equations, we obtain the rate of photons upconverted which is shown in figure \ref{fig:photonsUp}. We notice that using a 1 mm BBO crystal we should able to upconvert, and therefore measure, about 660 photons per second. This is well above the typical noise floor of a photomultiplier capable of single photon counting. Thus it should be possible to acquire a single point (for a given B-P delay) in $\sim$15 seconds with a signal-to-noise ratio of the order of $\sqrt{N}$=10$^2$. If 100 delays (2 ps/200 fs) are used to scan the entire time profile of the B pulse, its duration and skewness can be reliably reconstructed in 10 to 20 minutes.

\subsection{Timing resolution}
There are essentially three sources of uncertainty that limit our timing resolution. The first two are systematic uncertainties due to the laser jitter and the group velocity mismatch, which were discussed above to be lower than 150 fs and 200 fs, respectively. The third one arises from the statistical error on the skewness, which was calculated through a toy Monte Carlo simulation. In this simulation, a measurement is reproduced by a Gaussian histogram with $\sigma=10 \,ps$, 200 bins of width $200 \, fs$, and peak value of 10000 counts. Such a measurement would last, according to the estimation given in the preceding section, about 15 seconds per bin, and therefore 50 minutes in total. For the case of small skewness and large number of entries, it is possible to calculate the error on the skewness as

\begin{equation} \label{eq:skewnessError}
(\delta s)^{2} \approx \frac{\sum_{i}t_{i}^{2} N_{i}}{\sigma^{2} (\sum_{i}N_{i})^{2}}
\end{equation}

where $t_{i}$ and $N_{i}$ are the centers and the contents of the bins, respectively. From Figure \ref{fig:skewness}, we see that the skewness has an approximately linear dependence from the relative delay, with slope 0.36 obtained through linear fitting. Therefore, we have that the error on the relative delay is $\delta(\Delta z/\sigma_{z,nom}) \approx \delta s / 0.36$. From the Monte Carlo simulation, we obtained $\delta(\Delta z/\sigma_{z,nom}) \approx 0.003$, which correspond to an uncertainty in the timing of approximately 50 fs. Therefore, we have that all the uncertainties, both systematic ones and coming from statistics, lie below 200 fs. Considering all the uncertainties, we expect to be able to deliver a measure of the timing within an uncertainty of $1\%$ of the length of the radiating beam.

\subsection{Ultrafast LABM optical box}
\begin{figure}[h]
    \centering
    \includegraphics[width=1\textwidth]{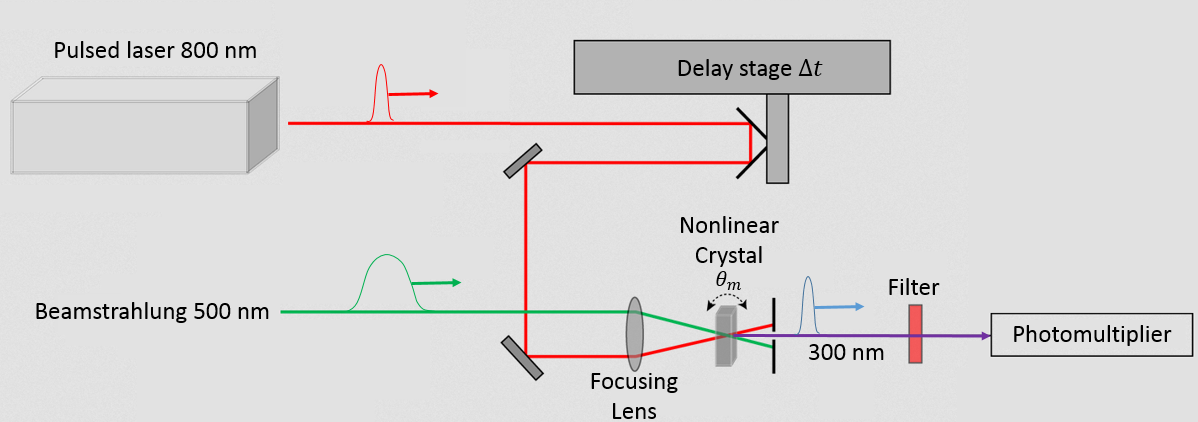}
    \caption{Setup of the Ultrafast LABM optical box.}
    \label{fig:upconversionSetup}
\end{figure}

The optical channel used to extract the beamstrahlung is already part of the instrumentation at SuperKEKB, therefore we only need to realize an optical box containing all the elements necessary to the upconversion technique. The Ultrafast LABM optical box will consist of a pulsed laser, a delay stage, some optical elements, and a detecting device, for example a photomultiplier. The setup is shown in figure \ref{fig:upconversionSetup}. 

The P pulse is much shorter then the B pulse, and it can be given a delay with a device called delay stage, which will be remotely controlled. The P pulse, with the given delay, interferes with the B pulse within the nonlinear crystal. Both pulses have to be focused of a small area of the crystal, in order to increase the efficiency. 

Suitably chosen mirrors, able to reflect only the visible portion of B radiation, will inject it into the optical box. Similarly, UV dielectric mirrors (or a filter) will be used after the nonlinear crystal, in order to eliminate photons which do not originate from upconversion, i.e. with a frequency lower than that expected for the upconverted photons. Finally, the photons are counted with a photomultiplier. Varying the relative delay of the pulses, it is possible to reconstruct the B pulse, and therefore have a measure of the timing of the beams and their length at the interaction point.

\section{Conclusion}
We have described a beam monitoring method that can be used to measure the timing and the length of the SuperKEKB beams at the interaction point. We expect to be able to measure the timing and the length of the beams with 1\% precision. 

The length of the beams can be measured with a resolution at the very least 5 times better than streak cameras. Beside this, the novelty of the method is that it allows establishing, with high accuracy, if the beams arrive at the IP simultaneously, or to fix them if they do not. Indeed, SuperKEKB beams will collide at a high crossing angle (83 mrad), introducing a novel possible way to lose luminosity when the bunches do not reach simultaneously the IP. 

It is noted that the method can be used also at synchrotron radiation sources, whenever a precise determination of the beam length is needed, by using a magnet short enough that the pulse time length is dominated by the beam length \cite{Beche:2004}. 

We have developed a completely original simulation of the collision of the beams in order to obtain the radiation field as a function of the time, which is ultimately what we aim to measure with our method. We have also presented a numerical calculation of the rate of photon upconversion, to show that the we have sufficient statistics to perform the measurement.

The experimental method of measurement involves an ultrafast pulsed laser, the use of a nonlinear crystal, and the phenomenon of photon upconversion. The technique involved has been thoroughly described along the paper, together with a description of the needed setup for a new Ultrafast LABM optical box. Basically, beamstrahlung is mixed with photons from the laser within a nonlinear crystal. A small fraction of the beamstrahlung photons in the visible range get upconverted to the ultraviolet and measured, allowing to reconstruct the beamstrahlung time profile. 

We are aware that luminosity is the first concern for SuperKEKB, and that every innovative beam monitoring system could be of vital importance for the success of the project. We believe that the monitoring system described in this paper is a valid candidate to be part of the SuperKEKB beam instrumentation.

\section*{Acknowledgment}
We would like to acknowledge Prof. Giovanni Bonvicini for many pieces of advice and fruitful discussions.


\end{document}